\documentclass[conference]{IEEEtran}

\usepackage[T1]{fontenc}
\usepackage[utf8]{inputenc}
\usepackage{array}
\usepackage[hidelinks]{hyperref}
\usepackage{url}
\usepackage{cite}
\usepackage{algorithmic}
\usepackage{graphicx}
\usepackage{textcomp}
\usepackage{enumitem}
\usepackage{tikz}
\usepackage{pgf-pie}
\usepackage{pgfplots}
\usepackage{color}
\usepackage{float}
\usepackage{lipsum}

\begin{document}
\title{Analysis of Service-oriented Modeling Approaches for Viewpoint-specific Model-driven Development of Microservice Architecture}

\author{\IEEEauthorblockN{Florian Rademacher, Sabine Sachweh}
	\IEEEauthorblockA{IDiAL Institute\\
		University of Applied Sciences and Arts Dortmund\\
		Otto-Hahn-Stra\ss{}e 23, 44227 Dortmund, Germany\\
		\{florian.rademacher,sabine.sachweh\}@fh-dortmund.de}
	\and
	\IEEEauthorblockN{Albert Z\"undorf}
	\IEEEauthorblockA{Department of Computer Science and Electrical Engineering\\
		University of Kassel\\
		Wilhelmsh\"oher Allee 73, 34121 Kassel, Germany\\
		zuendorf@uni-kassel.de}}

\maketitle

\begin{abstract}
Microservice Architecture (MSA) is a novel service-based architectural style for distributed software systems. Compared to Service-oriented Architecture (SOA), MSA puts a stronger focus on self-containment of services. Each microservice is responsible for realizing exactly one business or technological capability that is distinct from other services' capabilities. Additionally, on the implementation and operation level, microservices are self-contained in that they are developed, tested, deployed and operated independently from each other.

%However, next to these characteristics that distinguish MSA from SOA, both architectural styles rely on services as building blocks of distributed software architecture and hence face similar challenges regarding, e.g., service identification, composition and provisioning. Thereby, in contrast to MSA, SOA may rely on an extensive body of knowledge to tackle these challenges. Thus, due to both architectural styles being service-based, the question arises to what degree MSA might draw on existing findings of SOA research and practice.

Next to these characteristics that distinguish MSA from SOA, both architectural styles rely on services as building blocks of distributed software architecture and hence face similar challenges regarding, e.g., service identification, composition and provisioning. However, in contrast to MSA, SOA may rely on an extensive body of knowledge to tackle these challenges. Thus, due to both architectural styles being service-based, the question arises to what degree MSA might draw on existing findings of SOA research and practice.

In this paper we address this question in the field of Model-driven Development (MDD) for design and operation of service-based architectures. Therefore, we present an analysis of existing MDD approaches to SOA, which comprises the identification and semantic clustering of modeling concepts for SOA design and operation. For each concept cluster, the analysis assesses its applicability to MDD of MSA (MSA-MDD) and assigns it to a specific modeling viewpoint. The goal of the presented analysis is to provide a conceptual foundation for an MSA-MDD metamodel.
\end{abstract}

% For peerreview papers, this IEEEtran command inserts a page break and creates the second title. It will be ignored for other modes.
\IEEEpeerreviewmaketitle


\section{Introduction}
Microservice Architecture (MSA) \cite{Newman2015} is an architectural style for service-based software systems. It emphasizes high cohesion, loose coupling and self-containment of services. A \textit{microservice} denotes a component with well-defined interfaces that (i) provides a distinct, cohesive business or technological functionality to consumers; (ii) is autonomously developed and operated by one responsible team; (iii) comprises all technical artifacts for execution and deployment \cite{Nadareishvili2016,Francesco2017}. Expected benefits of MSA adoption comprise (i) better flexibility of software system adaptation, as services are independently deployable and replacement is simplified; (ii) increased service quality and safety, because of isolated testability, higher resilience and runtime scalability; (iii) increased development team productivity as the architecture's structure can be aligned to the team structure and each service being maintained by exactly one team \cite{Newman2015,Nadareishvili2016}.

%Compared to Service-oriented Architecture (SOA) \cite{Erl2005}, which is applied and studied for more than a decade, MSA is relatively young. It started to gain broad attention from practitioners and academia in 2014 and 2015, respectively \cite{Pahl2016}. However, with both architectural styles relying on services as building blocks for distributed software systems, they face similar challenges, e.g., in identifying, tailoring, composing and providing services \cite{Papazoglou2007}. Thus, it seems sensible to investigate findings from SOA research and practice with regard to their applicability to MSA. In this paper we focus on assessing the applicability of existing Model-driven Development (MDD) \cite{RodriguesDaSilva2015} approaches to SOA (SOA-MDD) in the context of MSA (MSA-MDD). Thereby, MDD is a research area whose adoption to SOA has been object to extensive research \cite{Ameller2015}.

Compared to Service-oriented Architecture (SOA) \cite{Erl2005}, which is applied and studied for more than a decade, MSA is relatively young. It started to gain broad attention from practitioners and academia in 2014 and 2015, respectively \cite{Pahl2016}. However, with both architectural styles relying on services as building blocks for distributed software systems, they face similar challenges, e.g., in identifying, tailoring, composing and providing services \cite{Papazoglou2007}. Thus, it seems sensible to investigate findings from SOA research and practice with regard to their applicability to MSA. In this paper we focus on assessing the applicability of existing Model-driven Development (MDD) \cite{RodriguesDaSilva2015} approaches to SOA (SOA-MDD) in the context of MSA (MSA-MDD). MDD is a research area whose adoption to SOA has been object to extensive research \cite{Ameller2015}.

%Specifically, the paper aims at providing the conceptual foundation for a metamodel, which enables MDD of microservice design and operation. It therefore presents an analysis of existing approaches to SOA modeling that yields a threefold contribution. First, our analysis surveys existing approaches to SOA modeling to identify and characterize modeling concepts related to service design and operation. Second, the analysis clusters semantically equivalent concepts of different modeling approaches and identifies clusters applicable to MSA-MDD. Third, it assigns applicable clusters to three \textit{modeling viewpoints} \cite{OMG2014} for MSA-MDD, i.e., Data, Service and Operation.

Specifically, the paper aims at providing the conceptual foundation of a metamodel to subsequently implement an MSA-MDD modeling language. Such a language could facilitate design and operation of MSA-based software systems, e.g., by code generation \cite{Rademacher2017}. It therefore presents an analysis of existing approaches to SOA modeling that yields a threefold contribution. First, our analysis surveys existing approaches to SOA modeling to identify and characterize modeling concepts related to service design and operation. Second, the analysis clusters semantically equivalent concepts of different modeling approaches and identifies clusters applicable to MSA-MDD. Third, it assigns applicable clusters to three modeling viewpoints \cite{OMG2014} for MSA-MDD, i.e., Data, Service and Operation.

The paper is organized as follows. Section~\ref{sec:soa-msa-mdd} presents differences between SOA and MSA relevant to service-based MDD. Section~\ref{sec:analysis} elucidates the analysis. Subsection~\ref{sub:analyzed-approaches} first introduces the considered SOA modeling approaches. Next, Subsections~\ref{sub:analysis-protocol}, \ref{sub:identified-concepts} and \ref{sub:viewpoint-clustering} present the analysis method and results, including SOA modeling concepts applicable to MSA-MDD. Section~\ref{sec:discussion} discusses these results. Section~\ref{sec:related-work} presents related work and Section~\ref{sec:conclusion} concludes the paper.

\section{Distinguishing Characteristics of Service-oriented and Microservice Architecture Relevant to Model-driven Development}\label{sec:soa-msa-mdd}
%MDD is a software engineering approach that considers models as means for abstracting the software to be built \cite{Combemale2017}, as well as first-class citizens in the engineering process \cite{RodriguesDaSilva2015}. In particular, the development of complex software systems benefits from employing MDD \cite{France2007,Whittle2014}. This is due to MDD providing \textit{modeling concepts} on the problem-level that abstract from implementation details. Thereby, sets of coherent concepts define \textit{metamodels}, i.e., abstract syntaxes for \textit{modeling languages}, which enable the expression of all models conforming to the underlying metamodel \cite{RodriguesDaSilva2015}. Another cornerstone of MDD is \textit{model transformation} \cite{Sendall2003}, e.g., the automatic generation of implementation artifacts from models.

MDD is a software engineering approach that considers models as means for abstracting the software to be built \cite{Combemale2017}, as well as first-class citizens in the engineering process \cite{RodriguesDaSilva2015}. In particular, the development of complex software systems benefits from employing MDD \cite{France2007,Whittle2014}. This is due to MDD providing \textit{modeling concepts} on the problem-level that abstract from implementation details. Sets of coherent concepts define \textit{metamodels}, i.e., abstract syntaxes for \textit{modeling languages}, which enable the expression of all models conforming to the underlying metamodel \cite{RodriguesDaSilva2015}. Another cornerstone of MDD is \textit{model transformation} \cite{Sendall2003}, e.g., the automatic generation of implementation artifacts from models.

%The application of MDD to SOA engineering is perceived to exhibit significant potential, because of the naturally high implementation complexity of distributed, service-based software systems \cite{Ameller2015,France2007}. Hence, lots of effort has been spent on SOA-MDD research \cite{Ameller2015} and practice-oriented standardization \cite{Kreger2009}. Thereby, SOA modeling puts a strong focus on generating code from models in the Development activity of software engineering \cite{Ameller2015}, which is commonly expected to increase developer productivity \cite{Hutchinson2011}. SOA-MDD metamodels and languages comprise \textit{modeling concepts} for services, interfaces, messages and ports, and are rather created from scratch than reused across the proposed MDD approaches \cite{Ameller2015}.

The application of MDD to SOA engineering is perceived to exhibit significant potential, because of the naturally high implementation complexity of distributed, service-based software systems \cite{Ameller2015,France2007}. Hence, lots of effort has been spent on SOA-MDD research \cite{Ameller2015} and practice-oriented standardization \cite{Kreger2009}. SOA modeling puts a strong focus on generating code from models in the Development activity of software engineering \cite{Ameller2015}, which is commonly expected to increase developer productivity \cite{Hutchinson2011}. SOA-MDD metamodels and languages comprise \textit{modeling concepts} for services, interfaces, messages and ports, and are rather created from scratch than reused across the proposed MDD approaches \cite{Ameller2015}.

While means for MSA-MDD might draw on the outlined knowledge about SOA-MDD, differences between SOA and MSA limit the applicability of existing SOA-MDD approaches to MSA \cite{Rademacher2017}. Thus, to substantiate the analysis of such approaches with the aim to identify SOA modeling concepts applicable to MDD of MSA design and operation (cf. Section~\ref{sec:analysis}), Table~\ref{tab:dcs-soa-msa} lists relevant \textit{distinguishing characteristics} (DCs) of SOA and MSA investigated in a previous work \cite{Rademacher2017}.

\begin{table*}
	\renewcommand{\arraystretch}{1.3}
	\caption{Selected Relevant Distinguishing Characteristics of Service-oriented and Microservice Architecture \cite{Rademacher2017}.}
	\label{tab:dcs-soa-msa}
	\centering
	\begin{tabular}{>{\centering\arraybackslash}p{0.3cm}|>{\centering\arraybackslash}p{1.8cm}|p{15cm}}
		\hline
		\bfseries \# & \bfseries Distinguishing Characteristic & \bfseries Peculiarity\\
		\hline\hline
		C1 & Service Granularity & MSA: Alignment of service functionality to a distinct business or technological capability. For business-related services, \textit{bounded contexts} \cite{Evans2004} for clustering and isolation of related domain concepts may be applied. SOA: No explicit guidance.\\
		\hline
		C2 & Interface Abstraction & MSA: Microservices and consumers typically have to use the same message formats/structures. SOA: Interaction of services with consumers using different message formats/structures is enabled by transformation capabilities of an Enterprise Service Bus (ESB).\\
		\hline
		C3 & Protocols & MSA: Promotes to apply at most two different communication protocols, one for one-to-one and one for one-to-many service interactions. SOA: ESBs may implement protocol transformations, enabling hypothetic support for an arbitrary amount of protocols.\\
		\hline
		C4 & Inter-service Interaction & MSA: For architecture-internal service interaction MSA prefers choreography over orchestration. SOA: SOA may equally apply both interaction patterns.\\
		\hline
		C5 & Extra-service Interaction & MSA: For architecture-external interactions \textit{API gateways} \cite{Balalaie2015}, i.e, rather simple fa\c{c}ades for abstracting services' endpoints and granularity, are employed. In particular, they do not implement sophisticated message or protocol transformation means like ESBs.\\
		\hline
		C6 & Application Scope & MSA: Mostly applied to (i) realize workflow-based applications with clear process flows; (ii) decompose monoliths with decreased scalability; (iii) realize web applications without generic message formats/structures or protocols. SOA: Typically applied in en\-ter\-prise-wide or cross-enterprise systems with heterogeneous message formats/structures, protocols or middleware technologies.\\
		\hline
		C7 & Practice Orientation & MSA: Higher perceived orientation towards practitioners due to (i) a reduced \textit{service taxonomy} \cite{Richards2015}; (ii) less complex implementation technologies, e.g., API gateways instead of ESBs; (iii) interoperable frameworks for implementation and provisioning of business-related and various infrastructural MSA components; (iv) focus on communication means perceived as being ``lightweight'', e.g., REST instead of SOAP \cite{Pautasso2008}; (v) teams' freedom of choice regarding service technologies, i.e., \textit{technology heterogeneity} \cite{Newman2015}.\\
		\hline
		C8 & Processes & MSA: Alignment of teams to features facilitates the application of \textit{container-based DevOps} \cite{Kang2016} within development processes.\\
		\hline
	\end{tabular}
\end{table*}

\section{Analysis of Service-oriented Modeling Approaches}\label{sec:analysis}
This section presents the protocol and results of our analysis of existing service-oriented modeling approaches. Therefore, Subsection~\ref{sub:analyzed-approaches} introduces the modeling approaches that were subject to our analysis. Following, Subsection~\ref{sub:analysis-protocol} presents the protocol we applied to perform our analysis, as well as general results of initial protocol steps. Subsection~\ref{sub:identified-concepts} presents the modeling concepts identified as being applicable to MSA-MDD and their semantic clustering. Subsection~\ref{sub:viewpoint-clustering} assigns applicable concepts to modeling viewpoints relevant to MSA-MDD, i.e., Data, Service and Operation.

\subsection{Analyzed Service-oriented Modeling Approaches}\label{sub:analyzed-approaches}
We considered ten existing SOA modeling approaches for our analysis. As our primary goal was to establish a conceptual foundation for subsequent deduction of an MSA-MDD metamodel that (i) targets two phases of MSA engineering, i.e., design and operation, and (ii) is preferably comprehensive, we selected modeling approaches that (i) aim at enabling holistic SOA modeling, i.e, define modeling concepts for multiple SOA engineering phases, and (ii) exhibit different degrees of abstraction and formality, i.e., comprise conceptual as well as practical elements. Specifically, our analysis comprises modeling approaches of the following types.

\paragraph*{Reference Models}
A \textit{reference model} is a conceptual framework that identifies relevant concepts of a given problem domain \cite{Kreger2009}. It further specifies relationships between concepts and is technology-independent.

\paragraph*{Reference Architectures}
\textit{Reference architectures} describe concepts of a problem domain with a focus on software architecture implementation \cite{Brown2012} and thus, compared to reference models, exhibit a lower degree of abstraction \cite{Kreger2009}.

\paragraph*{Modeling Languages and Profiles}
Modeling languages are based on metamodels that formally describe the structures of valid models expressed with the language (cf. Section~\ref{sec:soa-msa-mdd}). A metamodel may integrate a mechanism for tailoring or extending deduced modeling languages in \textit{modeling profiles} \cite{OMG2014}. Because modeling languages and profiles rely on a metamodel, they exhibit a high degree of formality, i.e., modeling concepts have well-defined structures and relationships, which enable applications to automatically process models \cite{Combemale2017}.

\paragraph*{Architecture Description Languages}
An \textit{architecture description language} (ADL) is a supportive modeling means for architecture-based software development \cite{Medvidovic2000}. In contrast to modeling languages and profiles, ADLs must comprise concepts for expressing architectural \textit{components}, \textit{connectors} and their \textit{configurations}. ADL-based configuration modeling also covers \textit{component composition} \cite{Medvidovic2000}, which is a crucial characteristic of service-based architectures \cite{Papazoglou2007}.

Table~\ref{tab:selected-approaches} lists the service-oriented modeling approaches, which we considered in our analysis. It also states, per approach, publication year, type with respect to abstraction and formality, foundational approaches if any and description.

\begin{table*}
	\renewcommand{\arraystretch}{1.3}
	\caption{Service-oriented Modeling Approaches Selected for Analysis and Identification of Modeling Concepts Applicable to MSA-MDD.}
	\label{tab:selected-approaches}
	\centering
	\begin{tabular}{c|p{4cm}|>{\centering\arraybackslash}p{0.5cm}|>{\centering\arraybackslash}p{1.2cm}|>{\centering\arraybackslash}p{1.5cm}|p{8cm}}
		\hline
		\bfseries \# & \bfseries Modeling Approach & \bfseries Year & \bfseries Type & \bfseries Foundational Approach & \bfseries Description\\
		\hline\hline
		A1 & Modeling and Design of Service-Oriented Architecture \cite{Stojanovic2004} & 2004 & Modeling Language & UML & The approach employs \textit{service components} as SOA building blocks. Interface-based design and UML are used to express capabilities as Business Service Components (BSCs). BSCs are composed from other BSCs and Application Service Components (ASCs) that implement fine-grained operations. Component interaction is contract-based.\\
		\hline
		A2 & A Modeling Framework for Service-Oriented Architecture \cite{Zhang2006} & 2006 & Modeling Language & 3C-modeling \cite{Szyperski2002} & Proposition of a metamodel with service components as first-class citizens of SOA modeling. The metamodel also comprises modeling concepts for expressing (i) required and provided ports to specify service interfaces; (ii) contracts; (iii) service choreographies.\\
		\hline
		A3 & Reference Model for Service Oriented Architecture \cite{MacKenzie2006} & 2006 & Reference Model & n/a & The reference model identifies essential SOA concepts and their relationships. Next to concepts for service, contract and interaction modeling, it considers execution contexts and visibility of services.\\
		\hline		
		A4 & A platform independent model for service oriented architectures \cite{Benguria2007} & 2007 & Modeling Language & MDA \cite{OMG2014} & The paper introduces the PIM4SOA approach. The underlying metamodel is structured on the basis of the four aspects (i) Information, i.e., modeling of information elements the other aspects rely on; (ii) Service, i.e., technology-independent description of business capabilities; (iii) Processes for modeling message-based service interactions; (iv) Quality of Service (QoS) addressing non-functional aspects.\\
		\hline
		A5 & A New Architecture Description Language for Service-Oriented Architecture \cite{Jia2007} & 2007 & ADL & XML & The presented SOADL language comprises concepts for modeling service interfaces, behavior, semantics and QoS-related aspects. Service composition is addressed by ``port bindings''.\\
		\hline
		A6 & Service-oriented Modeling Framework (SOMF) \cite{SOMF2011} & 2011 & Modeling Language & n/a & SOMF comprises a graphical modeling language for SOA-MDD. In addition to modeling concepts for services, interactions and compositions, it also considers modeling of cloud-based deployments as well as service, organizational and deployment boundaries.\\
		\hline
		A7 & SOA Reference Architecture \cite{OpenGroup2011} & 2011 & Reference Architecture & \cite{OpenGroupOntology2009},\newline approach~A3 & The approach clusters SOA modeling concepts in Architectural Building Blocks (ABBs). ABBs are assigned to five functional and four non-functional, cross-cutting layers. Relationships between ABBs and hence modeling concepts are expressed as layered ``interactions''.\\ 
		\hline
		A8 & Reference Architecture Foundation for Service Oriented Architecture \cite{Brown2012} & 2012& Reference Architecture & approach~A3 & Ascertainment of the reference model in approach~A3 with concept structures and additional concepts. The approach clusters concepts in Service Ecosystem, Realizing SOAs and Owning SOAs viewpoints.\\
		\hline
		A9 & Service oriented architecture Modeling Language (SoaML) Specification \cite{OMG2012} & 2012 & Modeling Profile & UML,\newline approach~A3 & The SoaML profile extends UML with concepts for SOA-MDD. It specifically provides means for sophisticated modeling of interfaces and interactions in the context of service-based software systems.\\
		\hline
		A10 & Topology and Orchestration Specification for Cloud Applications (TOSCA) \cite{Palma2013} & 2013 & Modeling Language & XML & TOSCA specifies a metamodel for expressing service deployment and operation. Despite not exclusively targeting SOA, we analyzed TOSCA to strengthen the consideration of operation-related modeling concepts. \\
		\hline
	\end{tabular}
\end{table*}

\subsection{Analysis Protocol and General Results}\label{sub:analysis-protocol}
In the following, we outline the protocol for our analysis of the approaches in Table~\ref{tab:selected-approaches} to identify service-oriented modeling concepts applicable to MSA-MDD and assign them to related modeling viewpoints. Furthermore, we present general results yielded by initial protocol steps.

The analysis protocol comprised the following steps:

\begin{enumerate}[label=\textbf{S.\arabic*},ref=S.\arabic*]
	\item \label{step:1concept-extraction} Identification and extraction of modeling concepts and concept-specific information from approach publications (cf. Table~\ref{tab:selected-approaches}).
	\item \label{step:2concept-characterization} Further characterization of modeling concepts per approach by surveying extracted concept-specific information. This step comprised three sub-steps.
	\begin{enumerate}[label=\textbf{S.2.\arabic*},ref=S.2.\arabic*]
		\item Identification of \textit{relationships} to other modeling concepts of the respective approach.
		\item Identification of \textit{structures}, i.e, concepts' properties not represented as relationships.
		\item Identification of formal \textit{constraints} that exceed multiplicity specifications for relationships and structures.
		\item Survey of textual concept descriptions that, next to formally expressed relationships, structures or constraints, express further semantics or characteristics of concepts.
	\end{enumerate}
	\item \label{step:3concept-reduction} Removal of concepts from the extracted set to which at least one of the following exclusion criteria applies:
	\begin{itemize}
		\item Concept is used in an SOA engineering phase other than design or operation.
		\item Concept enables advanced modeling of (i) architecture structure above service level; (ii) applicable concepts with lower abstraction level; (iii) governance; (iv) policies; (v) QoS; (vi) service-internal behavior.
		\item Concept lacks relationships, structure, constraints and its textual description is too generic or imprecise.
	\end{itemize}
	\item \label{step:4semantical-clustering} Identification and bundling of semantically equivalent remained concepts in \textit{concept clusters} across approaches.
	\item \label{step:5identification-applicable-concepts} Assessment of concept clusters' applicability to MSA-MDD on the basis of SOA and MSA DCs (cf. Table~\ref{tab:dcs-soa-msa}).
	\item \label{step:6viewpoint-clustering} Identification of modeling viewpoints for MSA-MDD and assignment of applicable concept clusters to them.
\end{enumerate}

The following paragraphs describe the executions of the initial steps~\ref{step:1concept-extraction} to ~\ref{step:3concept-reduction} and their general results. The executions and results of the main protocol steps~\ref{step:4semantical-clustering} to \ref{step:6viewpoint-clustering} are covered in more detail in Subsections~\ref{sub:identified-concepts} and \ref{sub:viewpoint-clustering}.

%In step~\ref{step:1concept-extraction} we performed a full reading of the approach publications listed in Table~\ref{tab:selected-approaches}. Thereby, we identified and extracted 447 modeling concepts. Next, step~\ref{step:2concept-characterization} yielded that of these concepts (i) 268 have relationships to others; (ii) 93 exhibit formal structure specifications; (iii) 15 comprise formal constraints not expressed in relationship or structure specifications. Table~\ref{tab:numbers-per-approach} breaks down these numbers per approach.

In step~\ref{step:1concept-extraction} we performed a full reading of the approach publications listed in Table~\ref{tab:selected-approaches} whereby we identified and extracted 434 modeling concepts. Next, step~\ref{step:2concept-characterization} yielded that of these concepts (i) 268 have relationships to others; (ii) 93 exhibit formal structure specifications; (iii) 15 comprise formal constraints not expressed in relationship or structure specifications. Table~\ref{tab:numbers-per-approach} breaks down these numbers per approach.

For reasons of space we do not present the raw results of steps~\ref{step:1concept-extraction} to \ref{step:3concept-reduction}, but provide them as supplemental material\footnote{Link to raw results of protocol steps~\ref{step:1concept-extraction} and \ref{step:2concept-characterization}: \url{https://fh.do/seaa2018-sm}}.

\begin{table}[H]
	\renewcommand{\arraystretch}{1.3}
	\caption{Numbers of Extracted Concepts and Characteristics per Approach after Execution of Protocol Step~\ref{step:2concept-characterization}.}
	\label{tab:numbers-per-approach}
	\centering
	\begin{tabular}{c|>{\centering\arraybackslash}p{1cm}|>{\centering\arraybackslash}p{1.5cm}|>{\centering\arraybackslash}p{1.2cm}|>{\centering\arraybackslash}p{1.3cm}}
		\hline
		\bfseries \# & \bfseries Concept Count & \bfseries ... with \newline Relationships & \bfseries ... with \newline Structures & \bfseries ... with \newline Constraints\\
		\hline\hline
		A1 & 10 & 7 & 2 & 0\\
		\hline
		A2 & 15 & 10 & 3 & 0\\
		\hline
		A3 & 25 & 21 & 2 & 0\\
		\hline		
		A4 & 35 & 35 & 4 & 0\\
		\hline
		A5 & 24 & 0 & 11 & 0\\
		\hline
		A6 & 45 & 0 & 0 & 0\\
		\hline		
		A7 & 12 & 10 & 0 & 0\\
		\hline
		A8 & 157 & 92 & 4 & 0\\ 
		\hline
		A9 & 34 & 27 & 18 & 15\\		
		\hline
		A10 & 77 & 48 & 49 & 0\\
		\hline\hline
		\bfseries $\Sigma$ & \bfseries 434 & \bfseries 250 & \bfseries 93 & \bfseries 15 \\
		\hline
	\end{tabular}
\end{table}

In step~\ref{step:3concept-reduction} we applied the mentioned exclusion criteria to each of the extracted and characterized modeling concepts. Hence, we filtered out concepts that do not support  design and operation, but other phases of the SOA engineering process. For example, these phases comprised requirements' elicitation with concepts like \texttt{Non-Function Requirement} of approach~A7 (cf. Table~\ref{tab:selected-approaches}) and business modeling with concepts like \texttt{Business-Goal Use Case} from approach~A1 or \texttt{Mo\-ti\-va\-tion\-El\-e\-ment} from approach~A9.

We further removed concepts from the extracted set for modeling of (i) architecture structure above service level, e.g., \texttt{Sub\-Ar\-chi\-tec\-ture} from approach~A5 or \texttt{In\-ter\-Cloud} from approach~A6; (ii) concepts at a lower abstraction level without being directly employable for SOA design or operation themselves, e.g., \texttt{Re\-la\-tion\-ship\-Type} from approach~A10; (iii) governance, e.g., \texttt{SOA Gov\-er\-nance} from approach~A8; (iv) policies, e.g., \texttt{Ser\-vice Pol\-icy} from approach~A3 or \texttt{Pol\-icy} from approach~A10; (v) QoS, e.g., \texttt{QoS\-Char\-ac\-ter\-is\-tic} form approach~A4; (vi) internal service behavior, e.g., \texttt{Be\-hav\-ior} from approach~A5.

The extracted concept set was additionally reduced by removing modeling concepts that did not exhibit relationships to other concepts of their defining approach, structures or constraints in combination with a too generic or imprecise textual description. The approaches~A3 and A8 comprise the majority of such concepts, e.g., \texttt{Real World Ef\-fect} and \texttt{Risk}. However, even publications of modeling languages, whose formality is expected to be high (cf. Subsection~\ref{sub:analyzed-approaches}), partially contain concept descriptions, which were insufficient for our analysis, e.g., \texttt{Context} from approach~A2.

\begin{figure}[H]
	\centering	
	\begin{tikzpicture}[scale=0.6, every node/.style={inner sep=0,outer sep=0}]
	\begin{axis}[
		ybar,
		bar width=.3cm,
		width=15.9cm,
		height=5cm,
		legend style={at={(0.68,0.9)}},
		symbolic x coords={A1, A2, A3, A4, A5, A6, A7, A8, A9, A10},
		xticklabels={
			A1\\\textcolor{red}{(50\%)}\\\textcolor{brown}{(40\%)}, 
			A2\\\textcolor{red}{(53\%)}\\\textcolor{brown}{(40\%)}, 
			A3\\\textcolor{red}{(36\%)}\\\textcolor{brown}{(36\%)}, 
			A4\\\textcolor{red}{(17\%)}\\\textcolor{brown}{(14\%)}, 
			A5\\\textcolor{red}{(46\%)}\\\textcolor{brown}{(46\%)}, 
			A6\\\textcolor{red}{(11\%)}\\\textcolor{brown}{(9\%)}, 
			A7\\\textcolor{red}{(25\%)}\\\textcolor{brown}{(25\%)}, 
			A8\\\textcolor{red}{(15\%)}\\\textcolor{brown}{(13\%)}, 
			A9\\\textcolor{red}{(50\%)}\\\textcolor{brown}{(38 \%)}, 
			A10\\\textcolor{red}{(16\%)}\\\textcolor{brown}{(6\%)}},
		xticklabel style={align=center},
		xtick=data,
		nodes near coords,
		nodes near coords align={vertical},
		ymin=0,ymax=180,
		legend cell align={left}
	]
	\addplot table[x=approach,y=conceptCount]{\selectedConceptShareData};
	\addplot table[x=approach,y=selectedCount]{\selectedConceptShareData};
	\addplot table[x=approach,y=applicableCount]{\selectedConceptShareData};
	\legend{Overall count of concepts after S.1, Count of remaining concepts after S.3, Count of applicable concepts after S.5 (cf. Subsection~\ref{sub:identified-concepts})}
	\end{axis}
	\end{tikzpicture}
	\caption{Comparison of concept counts after steps~\ref{step:1concept-extraction}, \ref{step:3concept-reduction} and \ref{step:5identification-applicable-concepts} of the analysis protocol (percentages rounded half up).}
	\label{fig:selected-concepts-share}
\end{figure}
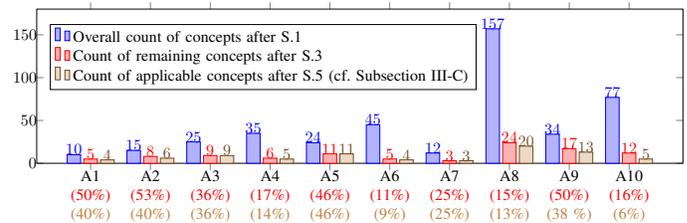

After finishing the execution of concept reduction step~\ref{step:3concept-reduction}, 100 concepts remained as inputs for protocol steps~\ref{step:4semantical-clustering} and \ref{step:5identification-applicable-concepts}. Figure~\ref{fig:selected-concepts-share} relates the count of those concepts to the overall concept count per approach. The first percentage under each approach is the share of remained in the overall concept count.

\pgfplotstableread[row sep=\\,col sep=&]{
	approach & conceptCount & selectedCount & applicableCount \\
	A1     & 10  & 5 & 4 \\
	A2     & 15 & 8 & 6 \\
	A3    & 25 & 9 & 9 \\
	A4   & 35 & 6 & 5 \\
	A5   & 24  & 11 & 11 \\
	A6   & 45  & 5 & 4 \\
	A7   & 12  & 3 & 3 \\
	A8   & 157  & 24 & 20 \\
	A9   & 34  & 17 & 13 \\
	A10   & 77  & 12 & 5 \\
}\selectedConceptShareData

\subsection{Semantic Clustering of Concepts and Assessment of Clusters' Applicability to Modeling for Microservice Architecture}\label{sub:identified-concepts}
%The execution of protocol step~\ref{step:4semantical-clustering} (cf. Subsection~\ref{sub:analysis-protocol}) included the encapsulation of modeling concepts, which remained after finishing protocol step~\ref{step:3concept-reduction}, in concept clusters across approaches. Thereby, the clustering was based on the semantics of these modeling concepts, as defined by their characteristics identified in step~\ref{step:2concept-characterization}. Next, each concept's applicability to MSA-MDD was assessed in step~\ref{step:5identification-applicable-concepts}. 

The execution of protocol step~\ref{step:4semantical-clustering} (cf. Subsection~\ref{sub:analysis-protocol}) included the encapsulation of modeling concepts, which remained after finishing protocol step~\ref{step:3concept-reduction}, in concept clusters across approaches. The clustering was based on the semantics of these modeling concepts, as defined by their characteristics identified in step~\ref{step:2concept-characterization}. Next, each concept's applicability to MSA-MDD was assessed in step~\ref{step:5identification-applicable-concepts}. 

%The results of steps~\ref{step:4semantical-clustering} and \ref{step:5identification-applicable-concepts} are shown in Tables~\ref{tab:fully-applicable-concepts} and \ref{tab:partially-applicable-concepts}. Table~\ref{tab:fully-applicable-concepts} lists concept clusters in which all concepts were assessed as being fully applicable to MSA-MDD. On the other hand, Table~\ref{tab:partially-applicable-concepts} includes all clusters that, next to fully applicable concepts, comprise at least one concept assessed as being partially applicable to MSA-MDD. Thereby, both tables exhibit an ``Approaches'' column, in which they state the modeling approaches whose concepts are part of the respective cluster. The column further states concepts' names that differ in their defining approaches from the respective cluster's name. Additionally, Table~\ref{tab:partially-applicable-concepts} differentiates partially (+) from fully (++) applicable modeling concepts, if both peculiarities of applicability occur within a concept cluster. Otherwise all concepts of a cluster in Table~\ref{tab:partially-applicable-concepts} were assessed as being partially applicable to MSA-MDD. For each concept assessed as being partially applicable, the ``Comments'' column of Table~\ref{tab:partially-applicable-concepts} contains a justification of the assessment that, if necessary, is based on the DCs of SOA and MSA (cf. Table~\ref{tab:dcs-soa-msa}).

The results of steps~\ref{step:4semantical-clustering} and \ref{step:5identification-applicable-concepts} are shown in Tables~\ref{tab:fully-applicable-concepts} and \ref{tab:partially-applicable-concepts}. Table~\ref{tab:fully-applicable-concepts} lists concept clusters in which all concepts were assessed as being fully applicable to MSA-MDD. On the other hand, Table~\ref{tab:partially-applicable-concepts} includes all clusters that, next to fully applicable concepts, comprise at least one concept assessed as being partially applicable to MSA-MDD. Both tables exhibit an ``Approaches'' column, in which they state the modeling approaches whose concepts are part of the respective cluster. The column further states concepts' names that differ in their defining approaches from the respective cluster's name. Additionally, Table~\ref{tab:partially-applicable-concepts} differentiates partially (+) from fully (++) applicable modeling concepts, if both peculiarities of applicability occur within a concept cluster. Otherwise all concepts of a cluster in Table~\ref{tab:partially-applicable-concepts} were assessed as being partially applicable to MSA-MDD. For each concept assessed as being partially applicable, the ``Comments'' column of Table~\ref{tab:partially-applicable-concepts} contains a justification of the assessment that, if necessary, is based on the DCs of SOA and MSA (cf. Table~\ref{tab:dcs-soa-msa}).

\begin{table*}
	\renewcommand{\arraystretch}{1.3}
	\caption{Concept Clusters Comprising only Modeling Concepts Assessed as Being Fully Applicable to MSA-MDD.}
	\label{tab:fully-applicable-concepts}
	\centering
	\begin{tabular}{c|p{3.2cm}|p{3.4cm}||c|p{3.2cm}|p{3.4cm}}
		\hline
		\bfseries \# & \bfseries Cluster & \bfseries Approaches (cf. Table~\ref{tab:selected-approaches}) & \bfseries \# & \bfseries Cluster & \bfseries Approaches \\
		\hline\hline		
		1 & Application Service & A5 & 17 & Message Exchange Pattern & A8 \\
		\hline
		2 & Architectural Service & A5 & 18 & Message Structure & A8: Structure \\
		\hline		
		3 & Artifact & A10: ArtifactTemplate, \newline ArtifactType & 19 & MessageType & A9 \\
		\hline
		4 & Atomic Service & A5: Atomic Service \newline A6: Analysis Atomic Service & 20 & Network Management & A8 \\
		\hline
		5 & Authentication & A8 & 21 & OutMessage & A5 \\
		\hline
		6 & Awareness & A3 & 22 & Plan & A10 \\
		\hline
		7 & Business Service & A5 & 23 & Port & A2, A9: Port \newline A5: Ports \\
		\hline
		8 & Categorization & A6: Service Typing Tag \newline A9: CategoryValues & 24 & Protocols & A8 \\
		\hline
		9 & Consumer & A6, A9: Consumer \newline A8: Service Consumer & 25 & Provider & A4: ServiceProvider \newline A8: Service Provider \newline A9: Provider \\
		\hline		
		10 & EndPoint & A4 & 26 & Reachability & A3: Reachability \newline A8: Service Reachability \\
		\hline
		11 & Execution Context & A3, A8: Execution Context \newline A9: ServicesArchitecture & 27 & RequirePort & A2: RequirePort \newline A9: Request \\
		\hline
		12 & FaultMessage & A5 & 28 & Security Management & A8 \\
		\hline
		13 & Infrastructure & A8 & 29 & Service Description & A3, A8 \\
		\hline		
		14 & InMessage & A5 & 30 & Specification & A1 \\
		\hline
		15 & Manageable Resource & A8 & 31 & Structure & A3, A8 \\
		\hline
		16 & Message Exchange & A8 & 32 & Visibility & A3 \\
		\hline
		\end{tabular}
\end{table*}

\begin{table*}
	\renewcommand{\arraystretch}{1.3}
	\caption{Concept Clusters Comprising at Least one Modeling Concept Assessed as Being Partially Applicable to MSA-MDD.}
	\label{tab:partially-applicable-concepts}
	\centering
	\begin{tabular}{c|p{2.8cm}|p{3.7cm}|p{9.5cm}}
		\hline
		\bfseries \# & \bfseries Cluster & \bfseries Approaches (cf. Table~\ref{tab:selected-approaches}) & \bfseries Comments on partial Applicability \\
		\hline\hline
		33 & Collaboration & A4, A9: Collaboration \newline A4, A9: CollaborationUse & Modeling of complex service collaborations with roles. MSA has a simpler view on service interaction (cf. DCs~C4 and C5 in Table~\ref{tab:dcs-soa-msa}). \\
		\hline
		34 & Composite Service & A5: Composite Service \newline A6: Analysis Composite Service \newline A8: Service & Sophisticated means for composing sets of fine-grained services with heterogeneous granularities to coarse-grained services. MSA facilitates composition as it proposes to align services to a self-contained, distinct functionality (cf. DC~C1). \\
		\hline
		35 & Enabling Technology & A7 & MSA relates technologies to services rather than architectural layers (cf. DC~C7). \\
		\hline
		36 & Information Model & A3 (++), A7 (++), A8 (+) & A8: In MSA-MDD, service models may not need to comprise abstracted message semantics for transforming exchange formats or structures (cf. DC~C2). \\
		\hline
		37 & Message & A4 (+), A5 (+), A8 (++) & A4: MSA exhibits a reduced application scope and service taxonomy not requiring role-based interaction modeling (cf. DCs~C6, C7). \newline A5: Microservices and consumers typically agree on the employed message formats/structures (cf. DCs~C2). Modeling of semantics may hence not be necessary. \\
		\hline
		38 & NodeTemplate & A10 & The \texttt{min\-In\-stances} and \texttt{max\-In\-stances} properties denote concept parts applicable to modeling of container-based microservice deployments (cf. DC~C8). \\
		\hline
		39 & NodeTypeImplementation & A10 & When associated with the concept, the semantics of the \texttt{De\-ploy\-ment\-Ar\-ti\-fact} and \texttt{Im\-ple\-men\-ta\-tion\-Ar\-ti\-fact} concepts are applicable to modeling of container-based microservice deployments (cf. DC~C8). \\
		\hline
		40 & Participant & A9 & Exhibits a high generality. Concept parts related to services are predominantly applicable to designing, e.g., microservice interfaces and contracts. \\		
		\hline
		41 & ProvidePort & A2: ProvidePort (++) \newline A9: Service (+) & A9: Abstract, technology-independent semantics and constraint models of service provisioning are not mandatory for MSA design and operation (cf. DCs~C2, C3). \\
		\hline
		42 & Service Component & A1: Service Component \newline A2: ServiceComponent & Both concepts provide, among others, means for sophisticated modeling of role-based service interaction, typically not applied by MSA (cf. DCs~C4, C5). \\
		\hline
		43 & ServiceContract & A1: Contract \newline A2, A9: ServiceContract \newline A3: Contract \& Policy & Concepts comprise conceptual or concrete parts for modeling complex service collaborations with respect to roles. MSA usually does not realize such complex collaborations (cf. DCs~C4, C5). \\
		\hline
		44 & Service Interface & A2: InterfaceDeclaration (+) \newline A9: Simple Interface (++) & A2: Service interfaces may not need abstract property or constraint descriptions for transforming message formats/structures or protocols (cf. DCs~C2, C3, C6).  \\
		\hline
		45 & Service Operation & A1 (+) \newline A3: Service (++) \newline A5, A8: Operation (++) & A1: Relationship to \texttt{Business-Goal Use Case} concept is not a mandatory prerequisite for MSA design and operation. \\
		\hline
		46 & Solution Building Block & A7 & Typically, MSA relates technology to services, not architectural layers (cf. DC~C7). \\
		\hline
		47 & Technical Assumptions & A8 & MSA-MDD may not need to consider modeling of physical limitations, e.g., flow speeds or disk access speeds (cf. DCs~C6, C7) \\
		\hline
		48 & Usage Management & A8 & Financial resource modeling on the service level may not be needed for MSA-MDD. \\
		\hline
	\end{tabular}
\end{table*}

%In total, we assessed 78 SOA modeling concepts as fully or partially applicable to MSA-MDD. Thereby, 54 modeling concepts were assessed as being fully applicable (69\%). Figure~\ref{fig:selected-concepts-share} also depicts the counts of applicable concepts per approach, as well as their shares in the overall concept count of the respective approach (second percentage under each approach).

In total, we assessed 80 SOA modeling concepts as fully or partially applicable to MSA-MDD with 54 modeling concepts been being assessed fully applicable (68\%). Figure~\ref{fig:selected-concepts-share} also depicts the counts of applicable concepts per approach, as well as their shares in the overall concept count of the respective approach (second percentage under each approach).

\subsection{Identification of Viewpoints for Microservice Architecture Modeling and Assignment of Applicable Concept Clusters}\label{sub:viewpoint-clustering}
The last protocol step~\ref{step:6viewpoint-clustering} of the analysis of the service-oriented modeling approaches comprised the identification of modeling viewpoints for MDD of MSA design and operation. A \textit{modeling viewpoint} provides a certain type of stakeholder with appropriate criteria to construct, select or present information about a system \cite{OMG2014}. It hence reduces the complexity practitioners of MDD have to deal with. As microservice teams usually apply DevOps \cite{Kang2016}, typical MSA stakeholder types are \textit{service developer} and \textit{operator}. Hence, we defined a viewpoint for each of these types. The \textit{Service viewpoint} for service developers comprises concept clusters from Tables~\ref{tab:fully-applicable-concepts} and \ref{tab:partially-applicable-concepts}, which focus modeling of microservices. The \textit{Operation viewpoint} encapsulates clusters for specifying aspects of service deployment and operation. It thus addresses service operators.

Next to these viewpoints for typical MSA stakeholder types, we also included a \textit{Data viewpoint}. Its addition was perceived to be sensible while executing protocol step~\ref{step:3concept-reduction}, because the analyzed approaches partially comprise explicit concepts for modeling information used by services. For example, \texttt{Struc\-ture} from approach~A3 (cluster~31 in Table~\ref{tab:fully-applicable-concepts}) enables expression of structural data types. While the number of concepts in the Data viewpoint is comparatively small, their semantics are extensive, e.g., as for \texttt{Information Model} from approaches~A3, A7 and A8 (cluster~36 in Table~\ref{tab:partially-applicable-concepts}).

Based on the described semantics and scopes of the viewpoints, we assigned the concept clusters from Tables~\ref{tab:fully-applicable-concepts} and \ref{tab:partially-applicable-concepts} to them as shown in Table~\ref{tab:viewpoints}. 

\begin{table}[H]
	\renewcommand{\arraystretch}{1.3}
	\caption{MSA-MDD Viewpoints With Assigned Concept Clusters And Their Defining Approaches.}
	\label{tab:viewpoints}
	\centering
	\begin{tabular}{c|p{4cm}|p{1.5cm}}
		\hline
		\bfseries Viewpoint & \bfseries Concept Clusters & \bfseries Approaches \\
		\hline\hline
		Data & 31, 36 & A3, A7, A8 \\
		\hline
		Service & 1, 2, 4, 7--9, 12, 14, 16--19, 21, 23, 27, 29, 33, 34, 37, 40--45 & A1--A6, A8, A9 \\
		\hline
		Operation & 3, 5, 6, 10, 11, 13, 15, 20, 22, 24--26, 28, 30, 32, 35, 38, 39, 46--48 & A1, A3, A4, A7--A10 \\
		\hline
	\end{tabular}
\end{table}

\begin{figure}[H]
	\centering
	\begin{tikzpicture}[scale=0.4, every node/.style={inner sep=0,outer sep=0}]
	\tikzstyle{every node}=[font=\scriptsize]
	\pie[text=legend, color={black!10, black!30, black!50}, explode={0, 0, 0}, after number=\%] {
		4/\scriptsize{Data (2 concept clusters)},
		52/\scriptsize{Service (25 concept clusters)},
		44/\scriptsize{Operation (21 concept clusters)}
	}
	\end{tikzpicture}
	\caption{Counts and shares of concept clusters assigned to viewpoints.}
	\label{fig:applicable-concepts-share}	
\end{figure}
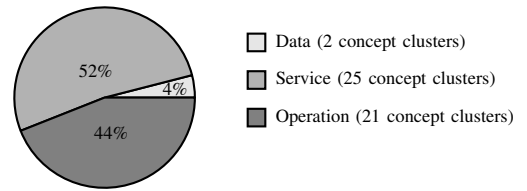

Figure~\ref{fig:applicable-concepts-share} shows for each viewpoint the count and share of assigned concept clusters.

The Data viewpoint comprises concept clusters for modeling data structures (cluster~31) and their integration in services' information models (36). 

The Service viewpoint's concept clusters enable modeling of (i) services and basic service roles (1, 2, 4, 7--9, 23, 27, 29, 40, 41); (ii) message types and structures (12, 14, 16--19, 21, 37, 42); (iii) interfaces and contracts (43--45); (iv) collaborations and compositions (33, 34).

The Operation viewpoint encapsulates concept clusters for modeling (i) service artifacts (3, 22); (ii) technical infrastructure (5, 11, 13, 15, 20, 28, 35, 38, 39, 46--48); (iii) service provisioning (6, 10, 24--26, 30, 32).

\section{Discussion}\label{sec:discussion}
Subsections~\ref{sub:metamodel-deduction} and \ref{sub:threats} discuss the impacts of the presented analysis results on subsequent metamodel deduction and elaborate on threats to the analysis's validity.

\subsection{Deduction of a Viewpoint-specific Metamodel for Microservice Architecture Modeling}\label{sub:metamodel-deduction}
%The results of our analysis provide the basis for deducing a viewpoint-specific metamodel for MSA-MDD with a focus on microservice design and operation. Thereby, the concept clusters shown in Tables~\ref{tab:fully-applicable-concepts} and \ref{tab:partially-applicable-concepts} may become initial concepts of the metamodel. Following, it is necessary to identify those characteristics of the applicable concepts (cf. Subsection~\ref{sub:analysis-protocol}) that may be adopted to the metamodel.

The results of our analysis provide the basis for deducing a viewpoint-specific metamodel for MSA-MDD with a focus on microservice design and operation. The concept clusters shown in Tables~\ref{tab:fully-applicable-concepts} and \ref{tab:partially-applicable-concepts} may become initial concepts of the metamodel. Furthermore, it would be necessary to identify those characteristics of the applicable concepts (cf. Subsection~\ref{sub:analysis-protocol}) that may be adopted to the metamodel.

Another aspect of the metamodel's deduction is the linkage of the identified viewpoints. For example, it is conceivable that the Service viewpoint needs to refer to the Data viewpoint so that message models may use data structures as types, or that Service and Operation viewpoint have to be associated to enable modeling of microservice deployment. Viewpoint linkage may be achieved by defining relationships between metamodel concepts of different viewpoints, either based on adopted concept characteristics or the introduction of intermediate concepts. 

Additionally, we expect the occurrence of contradictory overlaps in concept semantics when deducing the metamodel. In this case it has to be decided for each concept identified as being applicable to MSA-MDD in our analysis (cf. Section~\ref{sec:analysis}) how such inconsistencies may be dissolved on the metamodel level. It is further likely that the deduction process will reveal clusters that can be merged with others to comprehensively cover an aspect relevant to MSA design or operation in the metamodel. For example, the clusters \texttt{In\-Mes\-sage} and \texttt{Out\-Mes\-sage} (cf. Table~\ref{tab:fully-applicable-concepts}) may be represented by a property \texttt{di\-rec\-tion} of a superior \texttt{Mes\-sage} metamodel concept.

To cope with the mentioned challenges, we plan to deduce the metamodel in an iterative process, which continuously yields a more consistent version of the metamodel.

\subsection{Threats to Validity}\label{sub:threats}
Our analysis is affected by the following threats to validity.

\paragraph*{T1 Incomplete Selection of Modeling Approaches}
The analysis focused on surveying modeling approaches that take a holistic view on SOA-MDD (cf. Subsection~\ref{sub:analyzed-approaches}). However, it may be possible that we accidentally did not consider additional approaches with possible relevance to MSA design and operation. From our perspective, the impact of this threat is mitigated by the fact that Tables~\ref{tab:fully-applicable-concepts} and \ref{tab:partially-applicable-concepts} comprise clusters that directly correspond to the most recurrent metamodel concepts for modeling SOA design, i.e., \texttt{Service}, \texttt{Operation}, \texttt{Message} and \texttt{Port} \cite{Ameller2015}. In addition, all of these concepts occur in more than one analyzed approach, which makes us confident that they are central to service-oriented modeling in general. Considering the modeling of MSA operation, the threat's impact is likely to be lowered by the fact that the cluster count of the Operation viewpoint exhibits a scale similar to that of the Service viewpoint (cf. Figure~\ref{fig:applicable-concepts-share}).

\paragraph*{T2 Missing Applicable Concepts}
Despite the fact that we are confident to have captured essential modeling concepts applicable to MSA design and operation, we may have missed other relevant concepts of the analyzed approaches. As we executed reduction step~\ref{step:3concept-reduction} of our protocol (cf. Subsection~\ref{sub:analysis-protocol}) manually, it is possible that we accidentally filtered out modeling concepts with potential applicability to MSA-MDD. We tried to mitigate this threat's impact by double checking the results of the reduction step, as well as discussing and jointly deciding on edge cases. Furthermore, when deducing the metamodel, we plan to review further approaches, which specifically focus on a certain aspect of service-oriented modeling, to detail concept characteristics or identify additional concepts. This will apparently become necessary for the Data viewpoint, as it comprises comparatively few concept clusters (cf. Figure~\ref{fig:applicable-concepts-share}).

\paragraph*{T3 Lack of Practical Applicability}
MSA exhibits a high practice orientation (cf. DC~C7 in Table~\ref{tab:dcs-soa-msa}). However our analysis also considered modeling approaches with a rather theoretical focus, i.e., A3, A7 and A8. Hence, concept clusters may comprise elements with a comparatively low or even non-existent practical relevance. For now, we accept that this threat's impact may result in superfluous metamodel concepts. However, we plan to identify and remove such concepts irrelevant to practical MSA design and operation from the metamodel by surveying (i) related experience reports; (ii) related solution proposals; (iii) MSA developers and operators.

\section{Related Work}\label{sec:related-work}
In the following, we present work related to analyzing SOA modeling approaches and metamodeling for MSA-MDD.

%In 2013, Mohammadi et al. published a review of SOA modeling approaches for Enterprise Information Systems \cite{Mohammadi2013}. The review comprised seven modeling approaches and included the approaches~A3, A4, A6, A8 and A9 of our analysis (cf. Table~\ref{tab:selected-approaches}). The two remaining approaches are the SOA ontology \cite{OpenGroupOntology2009} on which approach~A7 is based and SOMA \cite{Arsanjani2008}, which leverages SoaML \cite{OMG2012} for service modeling, i.e., approach~A9 already included in our analysis. Thereby, the review is mainly focused on identifying and summarizing the major features of the considered modeling approaches. It does not analyze specific modeling concepts. Additionally, MSA in general and aspects of modeling SOA operation are not explicitly covered.

In 2013, Mohammadi et al. published a review of SOA modeling approaches for Enterprise Information Systems \cite{Mohammadi2013}. The review comprised seven modeling approaches and included the approaches~A3, A4, A6, A8 and A9 of our analysis (cf. Table~\ref{tab:selected-approaches}). The two remaining approaches are the SOA ontology \cite{OpenGroupOntology2009}, on which approach~A7 is based, and SOMA \cite{Arsanjani2008}, which leverages SoaML \cite{OMG2012} for service modeling, i.e., approach~A9 of our analysis. Furthermore, the review is mainly focused on identifying and summarizing the major features of the considered modeling approaches. It does not analyze specific modeling concepts. Additionally, MSA in general and aspects of modeling SOA operation are not explicitly covered.

Ameller et al. published results of a comprehensive mapping study on SOA-MDD based on 129 papers in 2015 \cite{Ameller2015}. Among others, one of their research goals was to investigate characteristics of SOA-MDD approaches (cf. Subsection~\ref{sub:analysis-protocol}) and the study reported on the seven most recurrent metamodel concepts in SOA-MDD. While considering a wide range of scientific publications for this purpose, the study did not involve an assessment of the concepts' applicability based on their semantics. Furthermore, MSA and modeling of SOA operation were out of the study's scope. However, papers from the study addressing modeling of SOA design may be surveyed for populating an initial MSA-MDD metamodel with further concepts or refining existing concepts. This would mitigate the impact of threat~T2 (cf. Subsection~\ref{sub:threats}).

%AjiL \cite{Sorgalla2017} is a language for graphical modeling of MSA-based software systems. It comprises concepts for basic modeling of (i) data structures; (ii) microservices and their interfaces; (iii) technical infrastructure. However, the metamodel lacks several \textit{essential service-oriented modeling concepts} \cite{Ameller2015} shown in Tables~\ref{tab:fully-applicable-concepts} and \ref{tab:partially-applicable-concepts}. For example, it does not comprise means for (i) modeling contracts and messages; (ii) specification of protocols and endpoints; (iii) expressing artifact and security technologies. Furthermore, AjiL does not define modeling viewpoints. Despite these shortcomings, AjiL exhibits a high practice orientation, as it is based on publicly available MSA implementations. Hence, we will review AjiL when deducing a metamodel on the basis of our analysis to raise its practical applicability and mitigate the impact of threat~T3.

AjiL \cite{Sorgalla2017} is a graphical language for MSA-MDD. It comprises basic concepts for modeling (i) data structures; (ii) microservices and their interfaces; (iii) technical infrastructure. However, the metamodel lacks several essential concepts of SOA modeling \cite{Ameller2015} shown in Tables~\ref{tab:fully-applicable-concepts} and \ref{tab:partially-applicable-concepts}. For example, it does not comprise concepts for specifying (i) service contracts and messages; (ii) protocols and endpoints; (iii) artifact and security technologies. Furthermore, AjiL does not define modeling viewpoints. Despite these shortcomings, AjiL exhibits a high practice orientation, as it is based on publicly available MSA implementations. Hence, we will review AjiL when deducing a metamodel on the basis of our analysis to raise its practical applicability and mitigate the impact of threat~T3.

\section{Conclusion and Future Work}\label{sec:conclusion}
This paper presented an analysis of existing SOA modeling approaches with the aim to identify modeling concepts applicable to viewpoint-specific MDD of MSA design and operation. Therefore, we first elucidated differences between SOA and MSA relevant to MDD (cf. Section~\ref{sec:soa-msa-mdd}). Next, we introduced the ten SOA modeling approaches (cf. Subsection~\ref{sub:analyzed-approaches}), which were analyzed following a rigorous protocol (cf. Subsection~\ref{sub:analysis-protocol}). Its first three steps yielded 100 modeling concepts possibly applicable to MDD of MSA design and operation. The applicability was then assessed considering SOA and MSA differences (cf. Subsection~\ref{sub:identified-concepts}). This resulted in 48 applicable concept clusters, being assigned to three MSA-MDD viewpoints (cf. Subsection~\ref{sub:viewpoint-clustering}). The analysis's discussion covered, among others, challenges of deducing a metamodel from the results (cf. Section~\ref{sec:discussion}).

In future works we plan to define a metamodel on the basis of the presented results. Our goal is to implement a modeling language for the metamodel to provide practitioners with generative means for microservice design and operation.

\bibliographystyle{IEEEtran}
\bibliography{IEEEabrv,literature}

\begin{thebibliography}{10}
\providecommand{\url}[1]{#1}
\csname url@samestyle\endcsname
\providecommand{\newblock}{\relax}
\providecommand{\bibinfo}[2]{#2}
\providecommand{\BIBentrySTDinterwordspacing}{\spaceskip=0pt\relax}
\providecommand{\BIBentryALTinterwordstretchfactor}{4}
\providecommand{\BIBentryALTinterwordspacing}{\spaceskip=\fontdimen2\font plus
\BIBentryALTinterwordstretchfactor\fontdimen3\font minus
  \fontdimen4\font\relax}
\providecommand{\BIBforeignlanguage}[2]{{%
\expandafter\ifx\csname l@#1\endcsname\relax
\typeout{** WARNING: IEEEtran.bst: No hyphenation pattern has been}%
\typeout{** loaded for the language `#1'. Using the pattern for}%
\typeout{** the default language instead.}%
\else
\language=\csname l@#1\endcsname
\fi
#2}}
\providecommand{\BIBdecl}{\relax}
\BIBdecl

\bibitem{Newman2015}
S.~Newman, \emph{Building Microservices}.\hskip 1em plus 0.5em minus
  0.4em\relax O'Reilly Media, 2015.

\bibitem{Nadareishvili2016}
I.~Nadareishvili, R.~Mitra, M.~Mclarty, and M.~Amundsen, \emph{Microservice
  Architecture}.\hskip 1em plus 0.5em minus 0.4em\relax O'Reilly Media, 2016.

\bibitem{Francesco2017}
P.~Di~Francesco, I.~Malavolta, and P.~Lago, ``Research on architecting
  microservices: Trends, focus, and potential for industrial adoption,'' in
  \emph{Proc. of the Int. Conf. on Software Architecture (ICSA)}.\hskip 1em
  plus 0.5em minus 0.4em\relax IEEE, 2017, pp. 21--30.

\bibitem{Erl2005}
T.~Erl, \emph{Service-Oriented Architecture (SOA) Concepts, Technology and
  Design}.\hskip 1em plus 0.5em minus 0.4em\relax Prentice Hall, 2005.

\bibitem{Pahl2016}
C.~Pahl and P.~Jamshidi, ``Microservices: A systematic mapping study,'' in
  \emph{Proc. of the 6th Int. Conf. on Cloud Computing and Services Science
  (CLOSER)}, 2016, pp. 137--146.

\bibitem{Papazoglou2007}
M.~P. Papazoglou and W.-J. Van Den~Heuvel, ``Service oriented architectures:
  approaches, technologies and research issues,'' \emph{The VLDB journal},
  vol.~16, no.~3, pp. 389--415, 2007.

\bibitem{RodriguesDaSilva2015}
A.~Rodrigues Da~Silva, ``Model-driven engineering: A survey supported by the
  unified conceptual model,'' \emph{Computer Languages, Systems and
  Structures}, vol.~43, pp. 139--155, 2015.

\bibitem{Ameller2015}
D.~Ameller, X.~Burgu\'{e}s, O.~Collell, D.~Costal, X.~Franch, and M.~P.
  Papazoglou, ``Development of service-oriented architectures using
  model-driven development: A mapping study,'' \emph{Information and Software
  Technology}, vol.~62, no.~1, pp. 42--66, 2015.

\bibitem{Rademacher2017}
F.~Rademacher, S.~Sachweh, and A.~Z{\"u}ndorf, ``Differences between
  model-driven development of service-oriented and microservice architecture,''
  in \emph{Proc. of the Int. Workshop on Architecting with MicroServices (AMS)
  co-located with ICSA}.\hskip 1em plus 0.5em minus 0.4em\relax IEEE, 2017.

\bibitem{OMG2014}
{Object Management Group}, \emph{Model Driven Architecture (MDA) Guide}, OMG
  Std., Rev. 2.0, 2014.

\bibitem{Combemale2017}
B.~Combemale, R.~B. France, J.-M. J\'{e}z\'{e}quel, B.~Rumpe, J.~Steel, and
  D.~Vojtisek, \emph{Engineering Modeling Languages}.\hskip 1em plus 0.5em
  minus 0.4em\relax CRC Press, 2017.

\bibitem{France2007}
R.~France and B.~Rumpe, ``Model-driven development of complex software: A
  research roadmap,'' \emph{Proc. of the 2007 Workshop on Future of Software
  Engineering (FOSE)}, 2007.

\bibitem{Whittle2014}
J.~Whittle, J.~Hutchinson, and M.~Rouncefield, ``The state of practice in
  model-driven engineering,'' \emph{IEEE software}, vol.~31, no.~3, pp. 79--85,
  2014.

\bibitem{Sendall2003}
S.~Sendall and W.~Kozaczynski, ``Model transformation: The heart and soul of
  model-driven software development,'' \emph{IEEE Software}, vol.~20, no.~5,
  pp. 42--45, 2003.

\bibitem{Kreger2009}
H.~Kreger and J.~Estefan, ``Navigating the soa open standards landscape around
  architecture,'' \emph{OASIS, OMG and The Open Group}, 2009.

\bibitem{Hutchinson2011}
J.~Hutchinson, J.~Whittle, M.~Rouncefield, and S.~Kristoffersen, ``Empirical
  assessment of mde in industry,'' in \emph{Proc. of the 33rd Int. Conf. on
  Software Engineering (ICSE)}.\hskip 1em plus 0.5em minus 0.4em\relax IEEE,
  2011, pp. 471--480.

\bibitem{Evans2004}
E.~Evans, \emph{Domain-Driven Design}.\hskip 1em plus 0.5em minus 0.4em\relax
  Addison-Wesley, 2004.

\bibitem{Balalaie2015}
A.~Balalaie, A.~Heydarnoori, and P.~Jamshidi, ``Migrating to cloud-native
  architectures using microservices: an experience report,'' in \emph{Workshop
  Proc. of the 4th Europ. Conf. on Service-Oriented and Cloud Computing
  (ESOCC)}.\hskip 1em plus 0.5em minus 0.4em\relax Springer, 2015, pp.
  201--215.

\bibitem{Richards2015}
M.~Richards, \emph{Microservices vs. Service-Oriented Architecture}.\hskip 1em
  plus 0.5em minus 0.4em\relax O'Reilly Media, 2015.

\bibitem{Pautasso2008}
C.~Pautasso, O.~Zimmermann, and F.~Leymann, ``Restful web services vs. "big"
  web services: Making the right architectural decision,'' in \emph{Proc. of
  the 17th Int. Conf. on World Wide Web}.\hskip 1em plus 0.5em minus
  0.4em\relax ACM, 2008, pp. 805--814.

\bibitem{Kang2016}
H.~Kang, M.~Le, and S.~Tao, ``Container and microservice driven design for
  cloud infrastructure devops,'' in \emph{Proc. of the Int. Conf. on Cloud
  Engineering (IC2E)}.\hskip 1em plus 0.5em minus 0.4em\relax IEEE, 2016, pp.
  202--211.

\bibitem{Brown2012}
P.~Brown, J.~A. Estefan, K.~Laskey, F.~G. McCabe, and D.~Thornton,
  \emph{Reference Architecture Foundation for Service Oriented Architecture
  Version 1.0}, OASIS Std., 2012.

\bibitem{Medvidovic2000}
N.~Medvidovic and R.~N. Taylor, ``A classification and comparison framework for
  software architecture description languages,'' \emph{IEEE Transactions on
  Software Engineering}, vol.~26, no.~1, pp. 70--93, 2000.

\bibitem{Stojanovic2004}
Z.~Stojanovic, A.~Dahanayake, and H.~Sol, ``Modeling and design of
  service-oriented architecture,'' in \emph{Proc. of the 2004 Int. Conf. on
  Systems, Man and Cybernetics (SMC)}, vol.~5.\hskip 1em plus 0.5em minus
  0.4em\relax IEEE, 2004, pp. 4147--4152.

\bibitem{Zhang2006}
T.~Zhang, S.~Ying, S.~Cao, and X.~Jia, ``A modeling framework for
  service-oriented architecture,'' in \emph{Proc. of the 6th Int. Conf. on
  Quality Software (QSIC)}.\hskip 1em plus 0.5em minus 0.4em\relax IEEE, 2006,
  pp. 219--226.

\bibitem{Szyperski2002}
C.~Szyperski, \emph{Component Software: Beyond Object-Oriented
  Programming}.\hskip 1em plus 0.5em minus 0.4em\relax Addison-Wesley, 2002.

\bibitem{MacKenzie2006}
C.~M. MacKenzie, K.~Laskey, F.~McCabe, P.~F. Brown, and R.~Metz,
  \emph{Reference Model for Service Oriented Architecture 1.0}, OASIS Std.,
  2006.

\bibitem{Benguria2007}
G.~Benguria, X.~Larrucea, B.~Elves{\ae}ter, T.~Neple, A.~Beardsmore, and
  M.~Friess, ``A platform independent model for service oriented
  architectures,'' in \emph{Enterprise Interoperability}.\hskip 1em plus 0.5em
  minus 0.4em\relax Springer, 2007, pp. 23--32.

\bibitem{Jia2007}
X.~Jia, S.~Ying, H.~Cao, and D.~Xie, ``A new architecture description language
  for service-oriented architecture,'' in \emph{Proceedings of the Sixth
  International Conference on Grid and Cooperative Computing (GCC)}.\hskip 1em
  plus 0.5em minus 0.4em\relax IEEE, 2007, pp. 96--103.

\bibitem{SOMF2011}
\emph{Service-oriented modeling framework (SOMF)}, Methodologies Corporation
  Std., Rev. 2.1, 2011.

\bibitem{OpenGroup2011}
\emph{SOA Reference Architecture}, The Open Group Std., 2011.

\bibitem{OpenGroupOntology2009}
{The Open Group}, \emph{The SOA Source Book}, 7th~ed.\hskip 1em plus 0.5em
  minus 0.4em\relax Van Haren, 2009.

\bibitem{OMG2012}
{Object Management Group}, ``Service oriented architecture modeling language
  ({SoaML}) specification version 1.0.1,'' 2012.

\bibitem{Palma2013}
D.~Palma and T.~Spatzier, \emph{Topology and Orchestration Specification for
  Cloud Applications Version 1.0}, OASIS Std., 2013.

\bibitem{Mohammadi2013}
M.~Mohammadi and M.~Mukhtar, ``A review of soa modeling approaches for
  enterprise information systems,'' \emph{Procedia Technology}, vol.~11, pp.
  794--800, 2013.

\bibitem{Arsanjani2008}
A.~Arsanjani, S.~Ghosh, A.~Allam, T.~Abdollah, S.~Ganapathy, and K.~Holley,
  ``Soma: A method for developing service-oriented solutions,'' \emph{IBM
  Systems Journal}, vol.~47, no.~3, pp. 377--396, 2008.

\bibitem{Sorgalla2017}
\BIBentryALTinterwordspacing
J.~Sorgalla, ``Ajil: A graphical modeling language for the development of
  microservice architectures,'' in \emph{Extended Abstracts of the
  Microservices 2017 Conference}, 2017. [Online]. Available:
  \url{http://www.conf-micro.services/papers/Sorgalla.pdf}
\BIBentrySTDinterwordspacing

\end{thebibliography}
\end{document}